# Multi-Layered Diagnostics for Smart Cities

Jungheum Park
School of Cybersecurity,
Korea University,
Seoul, South Korea
jungheumpark@korea.ac.kr

Hyunji Chung
School of Cybersecurity,
Korea University,
Seoul, South Korea
localchung@gmail.com

Joanna F. DeFranco
School of Graduate Professional
Studies, Pennsylvania State University,
Pennsylvania, U.S.A
jfd104@psu.edu

**ABSTRACT**

Smart cities use technology to improve traffic patterns, energy distribution, air quality and more. The elements of a smart city can also increase the convenience for its citizens, by integrating IT technology into many aspects of citizen interaction such as simplifying access to many of the city's services. The fields of healthcare, education, culture, and shopping can all be integrated into the core of a smart city to create an infrastructure that allows citizens to live more conveniently. Actual deployment cases exist in U.S., Europe, Singapore, and South Korea. With this environment, we need to think ahead about cybersecurity and prepare countermeasures as the cyberattacks in a smart city can threaten the lives of its citizens. In this paper, we examine smart city security threats from a multi-layered perspective, targeting representative elements that make up a smart city. A summary of attack scenarios and threat countermeasures are also described.

## 1 INTRODUCTION

A smart city aims to improve the quality of life of people by building an IT infrastructure that improves aspects of everyday life. As shown in **Figure 1**, smart cities have a multi-layer architecture in which each layer has multiple attack points to protect. Security and privacy are essential in a smart city environment as many end-point devices are connected to the Internet, communicate with each other, and accumulate data stored in a cloud. Cyberattacks can be very damaging on an individual level because smart cities are enabled by cyber-physical systems (CPS), which involve connecting devices and systems such as Internet of Things devices (IoT) [1].

Researchers have presented security and privacy challenges for intelligent healthcare, transportation, smart building, and smart energy [2]. However, there have been little studies on systematically addressing security with a multi-layer structure to include device/sensor level, application level, network level, and edge/cloud level from the viewpoint of security and privacy. In this paper, we will look at the elements and the attack points that potentially exist on multiple layers of a smart city (section 2). In addition, we will discuss the necessity of the security of a smart city by examining multiple scenarios (section 3). And finally, we will propose countermeasures to secure a smart city (section 4).

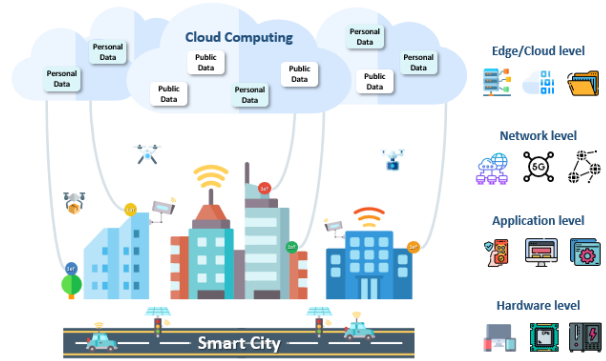

**Figure 1. Multi-layered Architecture of a Smart City**

## 2 ATTACK POINTS IN SMART CITY

**Table 1** explains the multi-level attack points for individual smart city elements: hardware (device/sensor), application, network, edge/cloud. Details of Each will be discussed in the following sub-sections.

### 2.1 Device/Sensor-level

Hardware chips of hub devices that control digital devices or sensors can be exploited by physical attacks. Smart sensors used by smart cities are often built into traffic lights and streetlights to manage parking system and reduce road congestion, but those wireless sensors can be vulnerable if they do not have security built in. In addition, there are cases where embedded systems used in traffic lights, healthcare devices, energy sensors, and home environments are hacked at the hardware level [3]. For example, vulnerabilities in Medronic's infrastructure were discovered that enabled an attacker to control an implanted pacemaker remotely due to tainted software updates [4].

### 2.2 Application-level

Each element consisting of a smart city uses web or mobile applications for the purpose of controlling devices/sensors and managing the data they generate. Thus, applications have frequently been the primary target of hacking with cross-site scripting (XSS) (e.g., injecting malicious scripts) and Man-in-the-Middle (MITM) (e.g., data transfer eavesdropping). It can be applied equally to a smart city. Through these attacks, any data transmitted in the network of the city can be intercepted or altered if there is no defense.

Table 1. Attack Points in Smart City

| Smart city elements | Device/Sensor-level | Application-level | Network-level | Edge/Cloud-level |
|---|---|---|---|---|
| ○ Smart transportation | ○ Smart car<br>○ Street sign<br>○ Smart traffic lights<br>○ Streetlights<br>○ Smart parking sensor | ○ Website<br>○ Application for desktop (Windows, macOS)<br>○ Application for mobile (iOS, Android) | ○ 4G, 5G<br>○ Wi-Fi<br>○ Zigbee<br>○ Z-Wave<br>○ Bluetooth<br>○ LoRa | ○ Public traffic<br>○ vehicle recording<br>○ signal control |
| ○ Smart healthcare | ○ Smart watch<br>○ Fitness tracker<br>○ Smart pacemaker<br>○ closed loop insulin delivery (artificial pancreas) | | | ○ Health data<br>○ GPS<br>○ Medical records |
| ○ Smart energy | ○ Air-conditioning<br>○ Heating sensor<br>○ Water leakage sensor<br>○ Light sensor<br>○ Temp & humidity sensor | | | ○ Customer location Temperature<br>○ Humidity<br>○ Power quality |
| ○ Smart building | ○ Smart thermostat<br>○ Smart camera<br>○ Smart speaker<br>○ Smart door lock<br>○ Smart baby monitor | | | ○ Human behavior<br>○ GPS, User voice<br>○ User interest<br>○ Movement<br>○ Photo/video |

Websites are at the application level and are indispensable to control the devices or systems that make up a smart city but unfortunately can be the first attack point for black hat hackers. For example, in 2017, an XSS attack occurred on eBay where malicious JavaScript was injected onto auction descriptions. This vulnerability was exploited to place malicious redirection code to a list of expensive vehicles [5].

## 2.3 Network-level

The devices, sensors, hubs, interfaces, and clouds that make up the smart city communicate organically with each other. For communication, there are various network protocols to enable short/local/wide-range communications such as Wi-Fi, Bluetooth, ZigBee, Z-Wave, LoRa, NB-IoT, WAVE, and 5G-eVTX [6]. Attacks that can occur at the network level include MITM attacks and packet tampering. There have been real cases in which data is captured and/or altered through these techniques. For example, MITM attacks have occurred to target Samsung refrigerators and other similar smart devices where hackers can infiltrate a network to steal the device owners Gmail credentials [7].

## 2.4 Edge/Cloud-level

The devices or sensors that make up the smart city generate digital data in real time and send it to the cloud, data analytics performed on this stored data can make smart cities smarter as it creates information that can add to improvements of the cities function. However, from a security point of view, a large amount of data stored remotely can be a threat. In a smart city, data generated from people, organizations, and government agencies are all stored remotely, so if related data is leaked, it can go beyond privacy infringement and be used for crime in the real world [8]. There are many cases where data stored in the cloud was leaked such as user personal information. Given cloud and edge data storages play a key role in smart cities, data breaches at the cloud level can cause serious secondary damage [9].

## 3 THREAT SCENARIOS

In this section, the reality of cyber threats in a smart city will be discussed. Threat scenarios are presented in **Table 2** in accordance to the attack techniques that can occur at each of the attack points [17]. **Table 2** maps the proposed threat scenarios to the MITRE's ATT&CK framework [16] in order to highlight prevalent smart city attack techniques and tactics. We describe the details of each scenario in the following sub-sections. In addition, **Table 3** maps the proposed threat scenarios to security parameters [19]. In other words, the security parameters address the confidentiality (protecting information from being accessed by unauthorized actors), integrity (ensuring data is not altered), and availability (data is accessible by authorized users) (CIA) model to the corresponding threat scenario.

Table 2. Threat Scenarios in Smart City

| Attack points in smart city | Related ATT&CK tactics | Representative ATT&CK techniques | Threat scenarios (sub-section #) | | | | |
|---|---|---|---|---|---|---|---|
| | | | (3.1) | (3.2) | (3.3) | (3.4) | (3.5) |
| ◦ Device/Sensor | ◦ Initial Access | ◦ Supply Chain Compromise | ✔ | ✔ | ✔ | ✔ | |
| | ◦ Persistence | ◦ Pre-OS Boot | ✔ | ✔ | ✔ | ✔ | |
| | ◦ Impact | ◦ Firmware Corruption | ✔ | ✔ | ✔ | ✔ | |
| ◦ Application | ◦ Credential Access | ◦ Brute Force | ✔ | ✔ | ✔ | | ✔ |
| | | ◦ Credentials from Password Stores | ✔ | ✔ | ✔ | | ✔ |
| | ◦ Impact | ◦ Data Destruction | ✔ | ✔ | ✔ | | |
| | | ◦ Data Encrypted for Impact | ✔ | ✔ | ✔ | | |
| | | ◦ Endpoint Denial of Service | ✔ | ✔ | ✔ | | |
| ◦ Network | ◦ Credential Access | ◦ Man-in-the-Middle | ✔ | ✔ | ✔ | ✔ | |
| | | ◦ Network Sniffing | | | | ✔ | |
| | ◦ Impact | ◦ Network Denial of Service | ✔ | ✔ | ✔ | | |
| | | ◦ Service Stop | ✔ | ✔ | ✔ | | |
| ◦ Edge/Cloud | ◦ Exfiltration | ◦ Exfiltration Over Other Network Medium | | | | | ✔ |
| | | ◦ Scheduled Transfer | | | | | ✔ |

Table 3. Proposed Scenarios and Security Parameters

| Threat scenarios | Security Parameters |
|---|---|
| ◦ Traffic chaos<br>◦ Medical ransomware<br>◦ Energy system hacking | ◦ Availability<br>◦ Integrity |
| ◦ Building attack | ◦ Authenticity<br>◦ Integrity |
| ◦ Security breach | ◦ Confidentiality |

## 3.1 Traffic chaos

Traffic paralysis due to traffic system hacking is the most prominent attack in smart cities where autonomous/automated driving and unmanned shuttle services, smart parking services, and other vehicle communications are routine. Hackers attacking autonomous/automated driving vehicles take control of the vehicles attempting to cause unexpected traffic chaos by forging and altering communication messages. For example, U.S. security company IO-Active successfully hacked using vulnerabilities in major urban traffic control system equipment, to demonstrate the fact that hackers can cause serious traffic jams or accidents [10].

At the device or sensor level, attacks such as supply chain compromise and firmware corruption are likely to occur. applications that control the transportation system can be attacked for a variety of reasons including credential access, data destruction, and endpoint denial of service. The devices that make up the smart transportation system are networked to each other, and thus attacks such as MITM, network denials of services and service stops can also pose threats to the entire smart city.

## 3.2 Medical ransomware

In smart cities, individual medical services and remote medical examinations using smart medical devices are becoming more prevalent. Thus, smart medical devices, associated applications, confidential medical information, big data systems, and medical data management centers are prime targets. If a medical device that is directly connected to life, such as a respirator or closed loop insulin pump is hacked, the patient's life is at risk. In 2018, a medical group's database was infected with malware, and about 16,000 prescriptions were leaked, including the local prime minister [11]. By exploiting these attacks, ransomware can be used to steal money from patients or their caregivers.

Attacks including supply chain compromise and firmware corruption are likely to be carried out at the medical equipment's device/sensor level. In order to gain access to the hospital's application that manages medical equipment or to the patient's application, hackers can attack to steal credentials and/or destruct important data for impact. Medical devices, medical institutions, and patients that make up smart medical systems are networked to each other, which can lead to frequent network-level attacks such as network denials of services and service stops. If this actually happens, it could pose a major threat to patients or medical institutions.

## 3.3 Energy system hacking

Cyberattacks aimed at sensors for energy production and consumption and related energy management systems can also cause tremendous damage to smart cities. Attacks against them are still active. In the Ukraine, power grids were hacked in 2015, causing 30 substations to stop working, and electricity supply was temporarily cut off. In April 2016, a power plant facility in Michigan, U.S., was infected with ransomware [12]. In particular, the attack was aimed at a programmable automatic control unit (PLC), which is a significant part of the operation of a factory and can have fatal consequences. If the PLC control of chemical factories is transferred to hackers, the neighborhood or the country is at risk – as this control produces a substance that is harmful to the human body.

At the hardware level of individual devices that constitute an energy management system, attacks on firmware of embedded systems can be major impacts. In addition, attackers targeting energy system related applications and networks can attempt to encrypt critical or sensitive data for financial gain as well as to perform denials of services like packet flooding and service down.

## 3.4 Building attack

When a house or company in a building is infected with a malware, it can be transmitted through communications of smart devices installed in the building. That is, if a hacker attacks an area of a building successfully, it can be said to have the potential to infect the entire building. People resting at home can be filmed by hacked smart TVs, and infected AI speakers can eavesdrop on personal conversations. In addition, video conferencing devices can record the content of meetings between employees in the company. This is a violation of privacy in that it monitors private activities.

At the hardware level, which is the device itself that makes up a smart home or office, attacks such as supply chain compromise and firmware corruption are likely to occur. In this environment, as in other scenarios, attacks on the network, for example, MITM, can occur because it is a networked environment. In addition, since smart homes or offices are relatively narrow spaces, attackers can steal data by network packet sniffing or prevent devices from operating properly through signal jamming.

## 3.5 Security breach

Hacking a data system in a smart city that uses big data to provide citizens with various convenience services is on a much larger scale of information leakage. Hackers' attacks on publicly sensitive data such as city administration, healthcare, and finance can shake the foundations of city operations. In 2017, a database of hospital network 'Atrium Health' was hacked in the U.S., and 2.65 million personal records were leaked [13]. This is a perfect example of the invasion of the privacy of citizens living in smart cities.

The security breach requires a key to access the space where the data is stored. Credentials such as username/password and access tokens would be the key, so attackers try to hijack it from target systems. There is also a possibility that attack techniques related to data capture at the edge/cloud level, exfiltration over other network medium, and scheduled transfer, may be applied. The leaked personal information can also be traded for money in illegal spaces such as the Dark Web. This could cause secondary damage, so leaking data is a very serious attack.

## 4 SECURING SMART CITY

In this section, we discuss potential strategies that will mitigate the risks discussed in the preceding sections. **Table 4** summarizes four countermeasures proposed in the following sub-sections.

## 4.1 Cybersecurity risk management

Urban planners and smart city governments need to proactively seek ways to make their cities, infrastructures safe from potential cybersecurity threats. In this context, smart cities must manage their security needs at multiple levels as explained in Section 3. It is necessary to analyze the security threats for each component constituting a smart city and establish protection measures for each layer according to the multiple levels described above.

A smart city should consider a plan of action using the NIST (National Institute of Standards and Technology) cybersecurity framework to improve the management of cybersecurity risks [18]. The framework provides detailed guidance to help develop a cybersecurity profile in order to prioritize and align activities with the city's risk tolerance and requirements.

Furthermore, in order to meet legal requirements such as the 'Personal Information Protection Act' and the 'Location Information Protection Act' in U.S., services must be developed by considering the information protection requirements from the initial phase of the business.

Table 4. Strategies for securing Smart Cities

| Countermeasures | Descriptions |
| --- | --- |
| Cybersecurity risk management | Urban planners and smart city governments need to proactively seek ways to make their cities, infrastructures safe from potential cyber threats. |
| Cyber patrol bot | A cyber patrol bot captures data on the overall state of a smart city system, including endpoint digital devices/sensors and network traffic. |
| Security and privacy label | Security and privacy labels for huge systems as well as single IoT products that make up a smart city should be provided with detailed descriptions of cybersecurity risks. |
| Nurturing talents | Suitable education programs can cultivate talented and skilled cybersecurity experts. |

## 4.2 Cyber patrol bot

Just as police patrol and monitor dangerous and vulnerable areas, it is necessary to monitor smart technology to detect and prevent cyberattacks. That is, a concept of a bot that can be used as a patrol system is needed to identify abnormalities of components consisting of a smart city. As an example, an implementation of the concept may capture data on the overall state of a smart city system, including endpoint digital devices/sensors and network traffic. This data can be analyzed to detect security vulnerabilities or potential security threats. Once any attack is detected, a broad range of actions should be operated.

The solution has several limitations. Cyber patrol bot will be effective in situations where various systems are interoperable. Thus, in order to implement this solution in the real world, it will only be possible for uniform systems. Moreover, it can be difficult to be interoperable with state-of-the-art systems even for legacy systems such as traffic management systems that are not state-of-the-art. Research is needed on how to integrate existing and new systems to create a monitoring system. As a countermeasure to the new environment, the solution from the policy/standard perspective and the technical solution will have to work complementarily.

## 4.3 Security and privacy label

Most consumer products are labeled with a description of the product for safety reasons. On groceries, the ingredients and processing plant information/location is provided to the consumer. Food nutrition labels were developed to decrease obesity by helping consumers purchase healthier food. Other goals of food nutrition labels include encouraging food companies to compete to produce healthier products without mandating specific nutritional requirements. Similar to this concept, a label for IoT devices, sensors, and systems that make up a smart city should be provided with detailed descriptions of the security and privacy risks. The concept of labeling a single product was proposed by Pardis Emami-Naeini et al. [14]. In a smart city, security and privacy labels for huge systems as well as single IoT products should be delivered to the administrator in the form of step-by-step instructions or notification windows. More specifically, as an example, to summarize the information to be included in the label:
- Security mechanisms
    - Security updates available until: Feb 27, 2020
    - Access control by: password, face-id, fingerprint
- Data practices
    - Collected sensor data types: text, audio, video
    - Data stored on: endpoint device, edge/cloud
    - Shared with: provider, manufacturer, authorized people

The security and privacy labels can help customers make decisions by informing them of potential risks that digital resources can pose to people when they purchase digital devices or build systems. Since there are many people in the city who do not have IT knowledge, labels should be created to a level that these people can easily understand.

## 4.4 Nurturing talented cybersecurity personnel

For the security of smart cities, an education program that can cultivate the best talent need to be established [17]. When developing smart city-related projects, local governments will face the same workforce issues familiar to everyone in the private sector: shortage of talent/workforce with greater cybersecurity skills. This only increases the security concerns inherent in these projects. To make our cities safer, stakeholders need to cooperate with schools and educational institutions to cultivate new workforces and systematically create programs to reeducate existing workforces. A complete CPS/IoT curriculum requires a skill set from existing engineering and computer science programs. To begin with, it is recommended to start with developing IoT/CPS elective courses in these major academic programs or at the very least, add CPS/IoT concepts to existing courses in those programs [15]. It may also be a good way for the country to retrain existing IT-related personnel or retirees who are experienced in it to take charge of protecting the city.

## 5  CONCLUSIONS

Smart cities, which were only in the imagination, are coming to reality. Several countries currently design and implement smart cities in major cities. As cities become smarter, citizens can live comfortably. But security researchers like us are responsible for keeping citizens safe from malicious attackers looking for vulnerabilities in smart city environments because cyber risks can pose a real threat as cities become increasingly instrumented and interconnected through information technologies. Because smart cities are very large-scale cities, there can be attacks on each element, device, and sensor that make up the city.

This work summarized vulnerabilities to the components that make up the city (e.g., smart transportation, smart healthcare, smart energy, and smart building) by dividing them into different attack points including hardware, applications, networks, and edge/cloud levels. We also presented five possible threat scenarios based on these attack points in consideration of the MITRE's ATT&CK techniques. Each threat scenario revolved around systems of infrastructure essential to the city, such as transportation, medical, and energy systems. It also dealt with possible threats in spaces such as home and office where people spend their days. Without ending with looking at the possible scenarios in smart cities, we jointly described solutions to make cities safe.

Based on the previously known cybersecurity framework, we proposed a concept of a bot that can detect threats on the system. Of course, we noted that there are interoperability issues in order to implement a cyber patrol bot in practice and the need for further research. Additional solutions include labeling food ingredients when selling food, recommending that digital devices or systems be sold with security labels when selling and building, as consumers can check ingredients when purchasing, and explaining the need for training and training professionals to keep cities safe in cybersecurity. It is natural for people living in smart cities to consider cybersecurity to live safely. We hope that our research will help stakeholders in smart cities including citizens, city authorities, governments, urban designers, service developers, security researchers, etc.

JUNGHEUM PARK is a research professor with the School of Cybersecurity, Korea University, Seoul, South Korea. His research interests include security, privacy, and digital forensics. Contact him at jungheumpark@korea.ac.kr.

HYUNJI CHUNG is a research professor with the School of Cybersecurity, Korea University, Seoul, South Korea. Her research interests include IoT forensics and smart city. Contact her at localchung@gmail.com (corresponding author).

JOANNA F. DeFRANCO is an associate professor of software engineering at Pennsylvania State University, University Park, Pennsylvania, 19355, USA.  Her research interests include software engineering Software Security, Distributed Networks, and Internet of Things.   Contact her at jfd104@psu.edu.